\documentclass[longauth]{aa}  
\usepackage{graphicx}
\usepackage{lscape}
\usepackage{longtable} 
\usepackage{pdflscape} 
\usepackage{gensymb}
\usepackage{txfonts}
\usepackage{subcaption}
\usepackage[dvipsnames]{xcolor}
\begin{document} 

\title{VLA cm-wave survey of young stellar objects in the Oph A cluster: constraining extreme UV- and X-ray-driven disk photo-evaporation}  

\subtitle{A pathfinder for Square Kilometre Array studies}

\titlerunning{VLA cm-wave survey of young stellar objects in the Oph A cluster}

   \author{
   A. Coutens\inst{1} 
   \and H. B. Liu\inst{2} 
   \and I. Jim{\'e}nez-Serra\inst{3}  
   \and T. L. Bourke\inst{4}   
   \and J. Forbrich\inst{5}  
   \and M. Hoare\inst{6}  
   \and L. Loinard\inst{7,8}   
   \and L. Testi\inst{2,9} 
   \and M. Audard\inst{10,11} 
   \and P. Caselli\inst{12} 
   \and A.~Chac\'on-Tanarro\inst{13}
   \and C. Codella\inst{9,14}
   \and J. Di Francesco\inst{15,16}
   \and F. Fontani\inst{9}  
   \and M. Hogerheijde\inst{17,18}
   \and A. Johansen\inst{19}
   \and D. Johnstone\inst{15,16}
   \and S. Maddison\inst{20}
   \and O. Pani\'c\inst{6}
   \and L. M. P\'erez\inst{21}
   \and L. Podio\inst{9}
   \and A. Punanova\inst{22}  
   \and J. M. C. Rawlings\inst{23}
   \and D. Semenov\inst{24,25}
   \and M. Tazzari\inst{26}  
   \and J. J. Tobin\inst{27}
   \and M. H. D. van der Wiel\inst{28}  
   \and H. J. van Langevelde\inst{29,16}  
   \and W. Vlemmings\inst{30}  
   \and C. Walsh\inst{6}  
   \and D. Wilner\inst{31}  
    }
     
 \institute{Laboratoire d'Astrophysique de Bordeaux, Univ. Bordeaux, CNRS, B18N, all\'ee Geoffroy Saint-Hilaire, 33615 Pessac, France \\
\email{audrey.coutens@u-bordeaux.fr} \and
European Southern Observatory (ESO), Karl-Schwarzschild-Str. 2, D-85748 Garching, Germany \and
School of Physics and Astronomy, Queen Mary University of London, Mile End Road, London E1 4NS, UK \and
SKA Organisation, Jodrell Bank Observatory, Lower Withington, Macclesfield, Cheshire SK11 9DL, UK \and
Centre for Astrophysics Research, University of Hertfordshire, College Lane, Hatfield AL10 9AB, UK \and
School of Physics and Astronomy, University of Leeds, Leeds LS2 9JT, UK \and
Instituto de Radioastronom\'{\i}a y Astrof\'{\i}sica, Universidad Nacional Aut\'onoma de M\'exico, Morelia 58089, M\'exico \and
Instituto de Astronom\'{\i}a, Universidad Nacional Aut\'onoma de Mexico, Apartado Postal 70-264, Ciudad de M\'exico 04510, M\'exico \and
INAF-Osservatorio Astrofisico di Arcetri, Largo Enrico Fermi 5, I-50125, Florence, Italy \and
Department of Astronomy, University of Geneva, Ch. des Maillettes 51, 1290 Versoix, Switzerland \and
Department of Astronomy, University of Geneva, Ch. d'Ecogia 16, 1290 Versoix, Switzerland \and
Max-Planck-Instit\"{u}t f\"{u}r extraterrestrische Physik, Giessenbachstrasse 1, 85748 Garching, Germany \and
Observatorio Astron\'omico Nacional (OAG-IGN), Alfonso XII 3, 28014, Madrid, Spain
\and Univ. Grenoble Alpes, Institut de Plan\'etologie et d'Astrophysique de Grenoble (IPAG), 38401 Grenoble, France
\and
NRC Herzberg Astronomy and Astrophysics, 5071 West Saanich Rd, Victoria, BC, V9E 2E7, Canada \and
Department of Physics and Astronomy, University of Victoria, Victoria, BC, V8P 5C2, Canada \and
Leiden Observatory, Leiden University, PO Box 9513, 2300 RA Leiden, The Netherlands \and
Anton Pannekoek Institute for Astronomy, University of Amsterdam, Science Park 904, 1098 XH Amsterdam, The Netherlands \and
Lund Observatory, Lund University, Box 43, 22100 Lund, Sweden \and
Centre for Astrophysics and Supercomputing, Swinburne University of Technology, Hawthorn, Victoria 3122, Australia \and
Departamento de Astronom\'ia, Universidad de Chile, Camino El Observatorio 1515, Las Condes, Santiago, Chile \and 
Ural Federal University, 620002, 19 Mira street, Yekaterinburg, Russia \and
Department of Physics and Astronomy, University College London, Gower St., London WC1E 6BT, UK \and
Department of Chemistry, Ludwig Maximilian University, Butenandtstr. 5-13,81377 M\"unchen, Germany \and
Max Planck Institute for Astronomy, K\"onigstuhl 17, D-69117, Heidelberg, Germany \and 
Institute of Astronomy, University of Cambridge, Madingley Road, CB3 0HA,  Cambridge, UK \and
NRAO, 520 Edgemont Road Charlottesville, VA 22903-2475 USA \and
ASTRON Netherlands Institute for Radio Astronomy, Oude Hoogeveensedijk 4, 7991 PD Dwingeloo, The Netherlands \and
Joint Institute for VLBI ERIC (JIVE), Oude Hoogeveensedijk 4, 7991 PD Dwingeloo, The Netherlands \and
Department of Space, Earth and Environment, Chalmers University of Technology, Onsala Space Observatory, 439 92 Onsala, Sweden \and
Harvard-Smithsonian Center for Astrophysics, 60 Garden Street, Cambridge, MA 02 138, USA
}

\date{Accepted XXX. Received YYY; in original form ZZZ}

\abstract
{Observations of young stellar objects (YSOs) in centimeter bands can probe the continuum emission from growing dust grains, ionized winds, and magnetospheric activity, which are intimately connected to the evolution of protoplanetary disks and the formation of planets. 
We have carried out sensitive continuum observations toward the Ophiuchus~A star-forming region using the Karl G. Jansky Very Large Array (VLA) at 10 GHz over a field-of-view of 6$'$ with a spatial resolution of $\theta_{\mbox{\scriptsize{maj}}}$\,$\times$\,$\theta_{\mbox{\scriptsize{min}}}$\,$\sim$\,0$\farcs$4\,$\times$\,0$\farcs$2. 
We achieved a 5 $\mu$Jy beam$^{-1}$ root-mean-square noise level at the center of our mosaic field of view. 
Among the eighteen sources we detected, sixteen are YSOs (three Class 0, five Class I, six Class II, and two Class III) and two are extragalactic candidates. 
We find that thermal dust emission generally contributes less that 30\% of the emission at 10 GHz. The radio emission is dominated by other types of emission such as gyro-synchrotron radiation from active magnetospheres, free-free emission from thermal jets, free-free emission from the outflowing photo-evaporated disk material, and/or synchrotron emission from accelerated cosmic-rays in jet or protostellar surface shocks. These different types of emission could not be clearly disentangled. 
Our non-detections towards Class II/III disks suggest that extreme UV-driven photoevaporation is insufficient to explain the disk dispersal, assuming that the contribution of UV photoevaporating stellar winds to radio flux does not evolve with time. The sensitivity of our data cannot exclude photoevaporation due to X-ray photons as an efficient mechanism for disk dispersal. Deeper surveys with the Square Kilometre Array will be able to provide strong constraints on disk photoevaporation.
}


\maketitle

\section{Introduction}

The first step towards forming the building blocks of planets occurs via grain growth in disks composed of dust and gas surrounding young stars \citep[e.g.,][]{Testi2014,Johansen2014}.
Thus, the time available for the formation of planets is limited by the lifetime of the disk.
After 10 Myr, the majority of disks disappear \citep[e.g.,][]{Haisch2005,Russell2006,Williams2011,Ribas2015}. Understanding the mechanisms leading to disk dispersal and the time-scales involved is crucial to characterizing  the environment in which planets form. 

The detection of transition disks where dust has been cleared in the inner regions \citep[e.g.,][]{Strom1989,Pascucci2016,vanderMarel2018,Ansdell2018} favored the development of theoretical models where disk dispersal occurs from the inside out (e.g., photoevaporation, grain growth, giant planet formation).
 In particular, models of disk dispersal through photoevaporation can successfully explain the inner hole sizes and accretion rates of a large number of transition disks \citep[e.g.,][]{Alexander2009,Owen2011,Owen2012,Ercolano2018}.
Given that radio observations trace ionized material, they could therefore provide useful constraints on different photoevaporation models  \citep{Pascucci2012,Macias2016}. Moreover, radio observations are also useful for tracing magnetospheric activity of the young stellar objects (YSOs) as well as grain growth process in disks \citep{Gudel2002,Forbrich2007,Forbrich2017,Choi2009,Guilloteau2011,Perez2012,Liu2014,Tazzari2016}.

The Ophiuchus\,A (hereafter Oph\,A) cluster is one of the nearest star-forming regions ($d\sim$137 pc, \citealt{Ortiz2017}).
Its proximity and richness in YSOs at a wide range of evolutionary stages \citep{Gutermuth2009} make this cluster an ideal laboratory for studying the evolution of YSO radio activity.
We present here the first results of new radio continuum observations of the Oph A region using the NRAO Karl G. Jansky Very Large Array (VLA) at 10 GHz, which achieve an unprecedented sensitivity (5 $\mu$Jy beam$^{-1}$ in the center of the field).
Section \ref{sect_obs} describes the observations and the data reduction. 
In Section \ref{sect_results}, we present the detected sources and analyze the nature of the continuum emission detected towards the YSOs. 
In Section \ref{sect_discussion}, we discuss the contribution of the Extreme-Ultraviolet (EUV) and X-ray photoevaporation in the dispersal of disks in Oph\,A, and prospects with the upcoming Square Kilometre Array (SKA). 

\section{Observations}
\label{sect_obs}

\begin{figure*}[!h]
   \begin{center}
      \hspace{-0.5cm}
      \includegraphics[width=14.5cm]{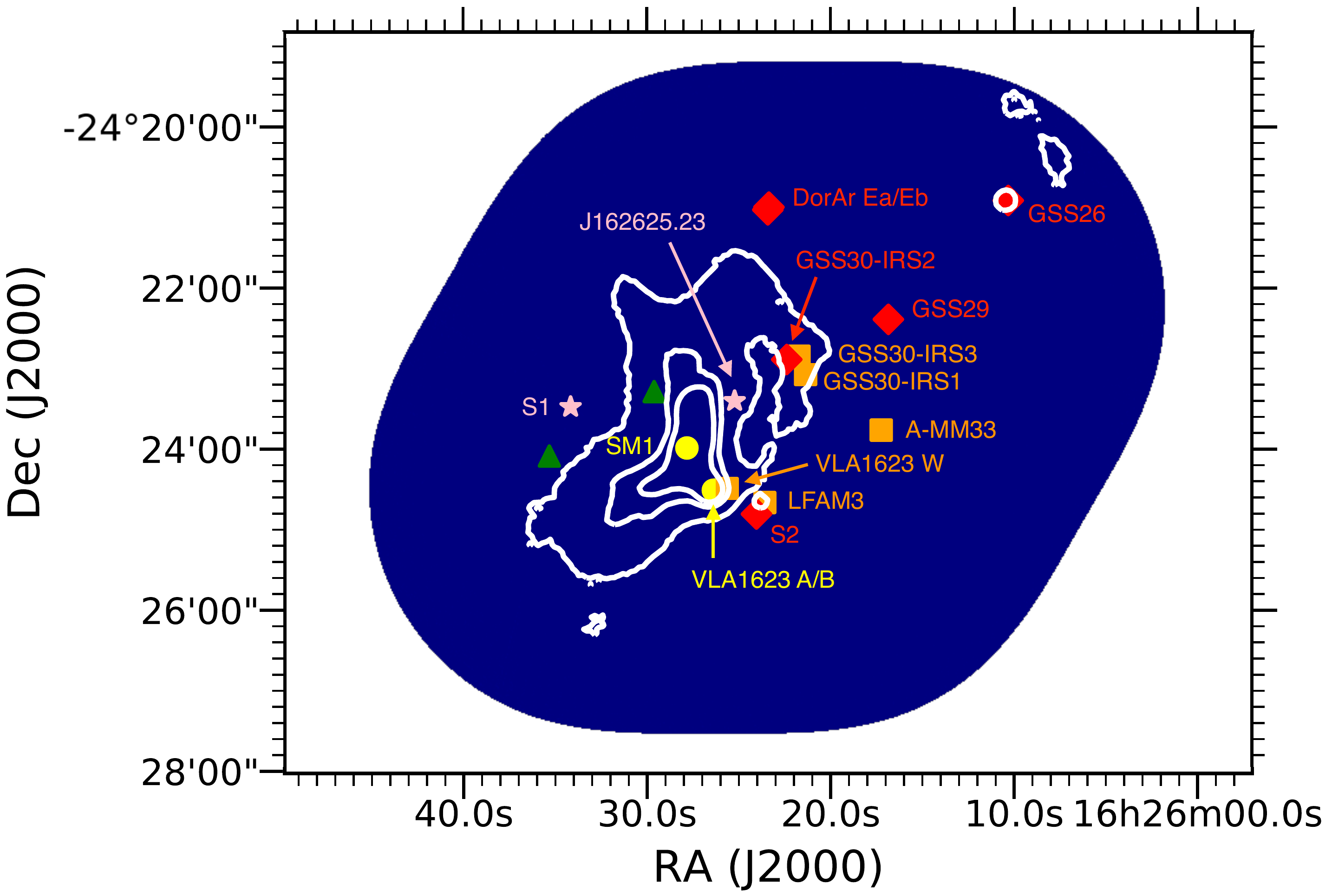} 
      \caption{Field of view covered by the VLA X band observations in blue. The position of the detected Class 0, I, II and III sources are indicated with yellow circles, orange squares, red diamonds, and pink stars, respectively. Sources VLA1623 and DoAr 24E are binary systems. The extragalactic candidates are indicated with green triangles. White contours represent 850 $\mu$m continuum observations from the JCMT Gould Belt Survey taken by SCUBA-2 \citep{Pattle2015,Kirk2018}.}
      \label{fig_global}
   \end{center}
\end{figure*}

We performed five epochs of mosaic observations towards the Oph\,A YSO cluster at X band (8.0--12.0 GHz) using the VLA (project code: 16B-259, PI: Audrey Coutens). All five epochs of observation (see Table \ref{tab:obstracks}) were carried out in the most extended, A array configuration, which provides a projected baseline range from 310 m to 34,300 m.
We used the 3-bit samplers and configured the correlator to have  4 GHz of continuous bandwidth
coverage centered on the sky frequency of 10 GHz divided into 32 contiguous spectral windows. 
The pointing centers of our observations are given in Table \ref{tab:pointings}. They are separated by 2.6$'$, while the primary beam FWHM is about 4.2$'$.
In each epoch of observation, the total on-source observing time for each pointing was 312 seconds.
The quasar J1625-2527 was observed approximately every 275 seconds for complex gain calibration.
We observed 3C286 as the absolute flux reference.
Jointly imaging these mosaic fields forms an approximately parallelogram-shaped mosaic field-of-view, of which the width and height are $\sim$6$'$. Figure \ref{fig_global} shows the observed field of view. 

We calibrated the data manually using the CASA\footnote{The Common Astronomy Software Applications software package, release 4.7.2 (\citealt{McMullin2007}).} software package, following standard data calibration procedures. 
For maximizing our sensitivity, we combined the data from all five epochs of observation. We ensured that highly variable sources did not affect the image quality or the results by also imaging the individual epochs separately (see Sect. \ref{sect_nature_remain_emission}). 
The imaging was done with Briggs robust = 2.0 weighting, gridder = `mosaic', specmode = `mfs', and nterms = 1. This setting was used to maximize S/N ratios. Using >1 nterms is not suitable for this project given that the sources are relatively faint. 
At the average observing frequency, we obtained a synthesized beam of $\theta_{\mbox{\scriptsize{maj}}}$\,$\times$\,$\theta_{\mbox{\scriptsize{min}}}$\,$\sim$\,0$\farcs$4\,$\times$\,0$\farcs$2 and a maximum detectable angular scale of $\sim$5$''$ (or $\sim$700 au).
After primary beam correction, we achieved a root-mean-square (RMS) noise level of $\sim$5 $\mu$Jy beam$^{-1}$ at the center of our mosaic field, which degrades to $\sim$28 $\mu$Jy beam$^{-1}$ toward the edges of the mosaic. 
The flux calibration uncertainty is expected to be about 5\%.

\begin{table}[!h]
   \caption{List of VLA observations}
   \vspace{-0.7cm}
   \label{tab:obstracks}
   \begin{center}
   {\footnotesize
      \begin{tabular}{@{}lc p{1.5cm}p{1.5cm} c@{}} 
      \hline\hline
      Epoch  &  Starting Time  &  Initial API rms & Projected baseline & $F^{\mbox{\scriptsize gain}}_{\mbox{\scriptsize 9.9 GHz}}$\\
             & (UTC)           &  ($^{\circ}$) & (m) & (Jy) \\
      \hline
      1 & 2016-12-02 21:31 & 11 & 310-34300 & 1.4 \\
      2 & 2016-12-05 21:18 & 6.0 & 460-34300 & 1.3 \\
      3 & 2017-01-06 18:05 & 13 & 325-32800 & 1.4 \\
      4 & 2017-01-14 18:40 & 13 & 310-34300 & 1.3\\
      5 & 2017-01-22 17:12 & 4.4 & 665-33100 & 1.3\\
      \hline
      \end{tabular}
   }
   \end{center}
   \vspace{-0.4cm}
\tablefoot{API refers to Atmospheric Phase Interferometer, which observes an 11.7 GHz beacon from a geostationary satellite with a 300 meters baseline. $F^{\mbox{\scriptsize gain}}_{\mbox{\scriptsize 9.9 GHz}}$ is the measured flux of the gain calibrator J1625-2527.}
\end{table}

\begin{table}[!h]
   \caption{List of the mosaic pointings}
   \vspace{-0.5cm}
   \label{tab:pointings}
   \begin{center}
   {\footnotesize
      \begin{tabular}{lll}
      \hline\hline
         Name  &  R.A.  & Decl. \\
               & (J2000)   & (J2000)  \\
         \hline
         X1 & 16$^{\mbox{\tiny h}}$26$^{\mbox{\tiny m}}$32$^{\mbox{\tiny s}}$.00 & 
              -24$^{\circ}$24$'$30$\farcs$0 \\
         X2 & 16$^{\mbox{\tiny h}}$26$^{\mbox{\tiny m}}$20$^{\mbox{\tiny s}}$.62 & 
              -24$^{\circ}$24$'$30$\farcs$0 \\
         X3 & 16$^{\mbox{\tiny h}}$26$^{\mbox{\tiny m}}$26$^{\mbox{\tiny s}}$.31 & 
              -24$^{\circ}$22$'$15$\farcs$0 \\
         X4 & 16$^{\mbox{\tiny h}}$26$^{\mbox{\tiny m}}$14$^{\mbox{\tiny s}}$.93 & 
              -24$^{\circ}$22$'$15$\farcs$0 \\
      \hline
      \end{tabular}
   }
   \end{center}
   \vspace{-0.4cm}
\end{table}


\section{Results}
\label{sect_results}

\subsection{Source census and comparison with other surveys}
\label{sect_census}

In total, we detect 18 sources above 5$\sigma$ in our mosaic field of view. The fluxes of the detected sources were measured by performing two-dimensional Gaussian fits, using the {\tt imfit} task of CASA.
The derived fluxes and coordinates can be found in Table \ref{table_catalog}, where the names of the sources J{\it hhmmss.ss-ddmmss.s} are based on the coordinates of peak intensity obtained with the fitting procedure. The position uncertainties are typically about a few tens of mas. 
Table \ref{table_catalog} also lists the more commonly used names of these sources. When the source structure was too complex to be fitted with this method or the results of the fit were too uncertain, we measured the flux by integrating over a circular area around the source with CASA.
Table \ref{table_imfit} summarizes the sizes measured with the Gaussian fit after deconvolution from the beam. 

\begin{figure*}[!h]
\begin{center}
\includegraphics[width=17.5cm]{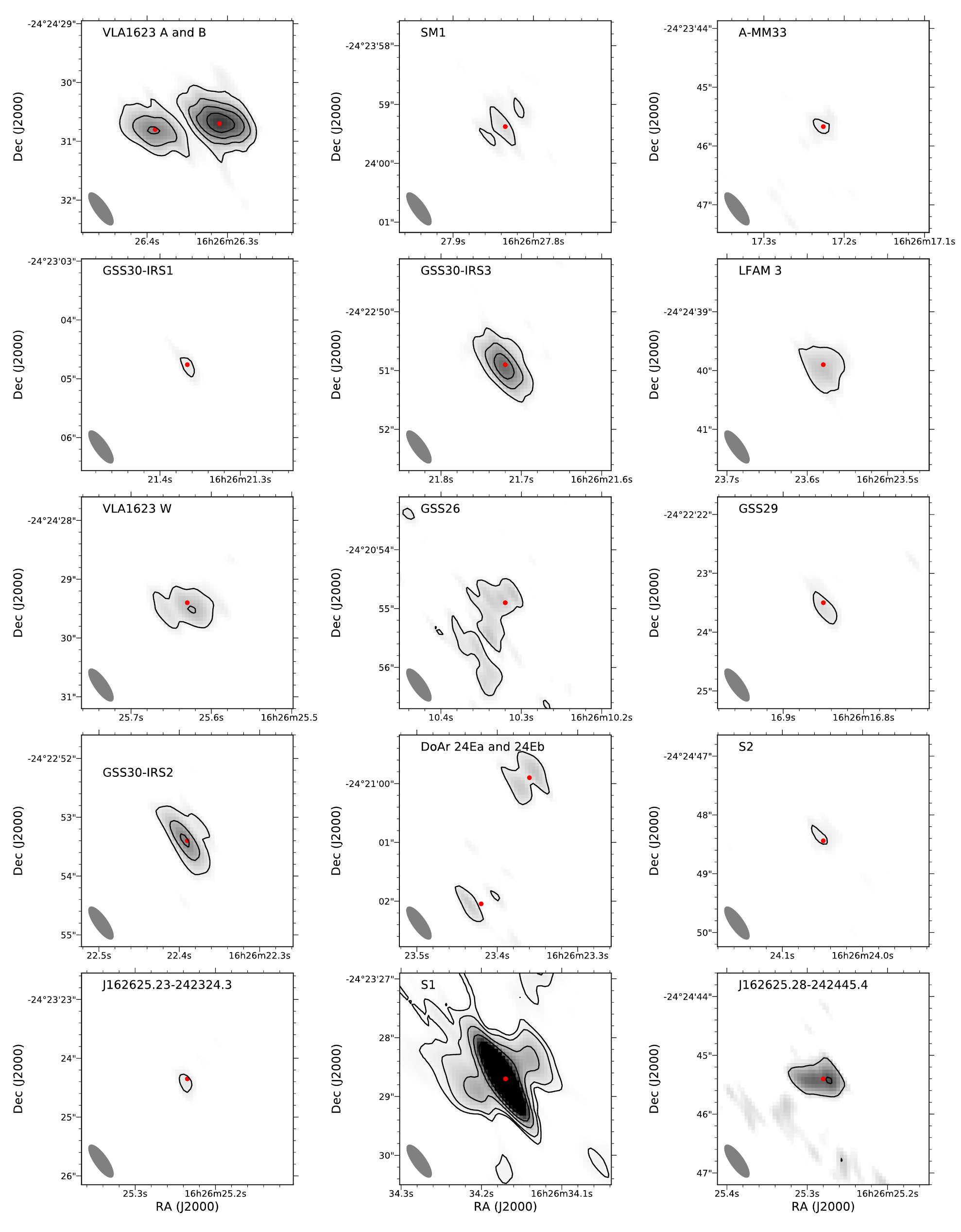} 
\caption{Continuum observations of the detected young stellar objects with the VLA in band X. For all sources except S1, the contours start from 5$\sigma$ with a step of 5$\sigma$. For S1, the contours are 10$\sigma$, 20$\sigma$, 50$\sigma$, 100$\sigma$, and 200$\sigma$. The greyscale images start at 3$\sigma$. The red dot corresponds to the coordinates used to name the sources in Table \ref{table_catalog}. The last map (J162625.28-242445.4) is for epoch 3 only. }
\label{fig_indiv_maps}
\end{center}
\end{figure*}

We compared our detections with the list of YSOs present in our field based on the photometric and spectroscopic surveys presented in \citet{Wilking2008}, \citet{Jorgensen2008}, \citet{Hsieh2013}, and \citet{Dzib2013}. Previously \citet{Dzib2013} carried out large-scale observations of the Ophiuchus region with the VLA at 4.5 GHz and 7.5 GHz with a resolution of 1$\arcsec$. 
\citet{Wilking2008} used X-ray and infrared photometric surveys as well as spectroscopic surveys of the L1688 cloud to list all the association members present in the Two Micron All-Sky Survey (2MASS) catalog. They also classified the sources according to their respective spectral energy distributions (SEDs) built from the Spitzer Cores to Disks (c2d) survey.
\citet{Hsieh2013} compiled another list based on the c2d Legacy Project after developing a new method to identify fainter YSOs based on analyzing multi-dimensional magnitude space. Finally, \citet{Jorgensen2008} identified the more deeply embedded YSOs by jointly analyzing Spitzer and JCMT/SCUBA data. 
Our 18 detected sources have all been found in at least one earlier catalog or study. Specifically, 16 of our 18 radio detections are associated with YSOs, while the remaining two are probably extragalactic sources \citep{Dzib2013}. Individual images of our detected YSOs are provided in Figure \ref{fig_indiv_maps}. Sidelobes are visible for some of these sources (S1, SM1). 
In total, we detect 11 YSO candidates listed in the catalog of \citet{Wilking2008}, while the remaining 19 YSOs in that catalog are undetected (see label "b" in Table \ref{table_catalog}).
Also, we detected nine of the sources listed by \citet[][see label "c" in Table \ref{table_catalog}]{Hsieh2013}. 
Finally, we detected five of the young sources listed in \citet{Jorgensen2008}, while two others (162614.63-242507.5 and 162625.49-242301.6) are undetected.

Compared to the previous VLA survey at 4.5 GHz and 7.5 GHz  by \citet[][see Table \ref{table_catalog}]{Dzib2013}, we detect seven additional radio sources, namely J162627.83-242359.4 (SM1, \#3 in Table \ref{table_catalog}), J162617.23-242345.7 (A-MM33, \#4), J162621.36-242304.7 (GSS30-IRS1, \#5), J162623.36-242059.9 (DoAr 24Ea, \#19), J162623.42-242102.0 (DoAr 24Eb, \#20), J162624.04-242448.5 (S2, \#21), and J162625.23-242324.3 (\#30). All of these are young stellar objects.
Three sources reported in \citet{Dzib2013} are undetected in our observations. These three sources are extragalactic (EG) candidates. A possible explanation for these non-detections is that they have a negative spectral index. Hence, the observations at 4.5 GHz and 7.5 GHz by \citet{Dzib2013} could be more sensitive to this type of targets because of their higher brightness at lower frequency. Another  explanation would be that these sources are variable. We will now comment briefly on some of the individual young stellar objects.

J162627.83-242359.3 (also known as SM1, \#3) was previously classified as a prestellar core (see \citealt{Motte1998}). 
It was, however, detected at 5 GHz with the VLA at an angular resolution of $\sim$10$\arcsec$ (measured peak fluxes of 130 -- 200 $\mu$Jy beam$^{-1}$; \citealt{Leous1991,Gagne2004}), although, in the first study, the source appears slightly offset by 3$\arcsec$. More recent ALMA observations suggest that SM1 is actually protostellar and that it hosts a warm ($\sim$30--50 K) accretion disk or pseudo-disk \citep{Friesen2014,Kirk2017,Friesen2018}. 

The source J162623.42-242101.9 (known as DoAr\,24Eb, \#20) is the companion of the protostar J162623.36-242059.9 (DoAr\,24Ea, \#19), also detected in our dataset (see Figure \ref{fig_indiv_maps}). 
These two sources are assumed to be at a similar evolutionary stage, although more data are needed to confirm this hypothesis \citep{Kruger2012}. 

The source J162634.17-242328.7 (S1, \#32) was suggested to be a binary separated by 20-30 mas (see discussion in \citealt{Ortiz2017}). 
Our VLA X band image does not spatially resolve the individual binary components. We note that the secondary component was not detected in the most recent epochs covered by \citet{Ortiz2017}.

For the sources that we did not detect, we evaluated the 3$\sigma$ upper limits, which vary across the mosaic field due to primary beam attenuation (see Table \ref{table_catalog}).
For 3\,$\sigma$ RMS levels as low as $\sim$15 $\mu$Jy beam$^{-1}$, the detection statistics at 10 GHz in this region are 3/3 for Class 0 sources (100\%), 5/8 for Class I YSOs (63\%), 6/16 Class II sources (38\%), and 2/5 Class III objects (40\%). 

\begin{figure}[!h]
   \begin{tabular}{ p{10cm} }
     \includegraphics[width=9.5cm]{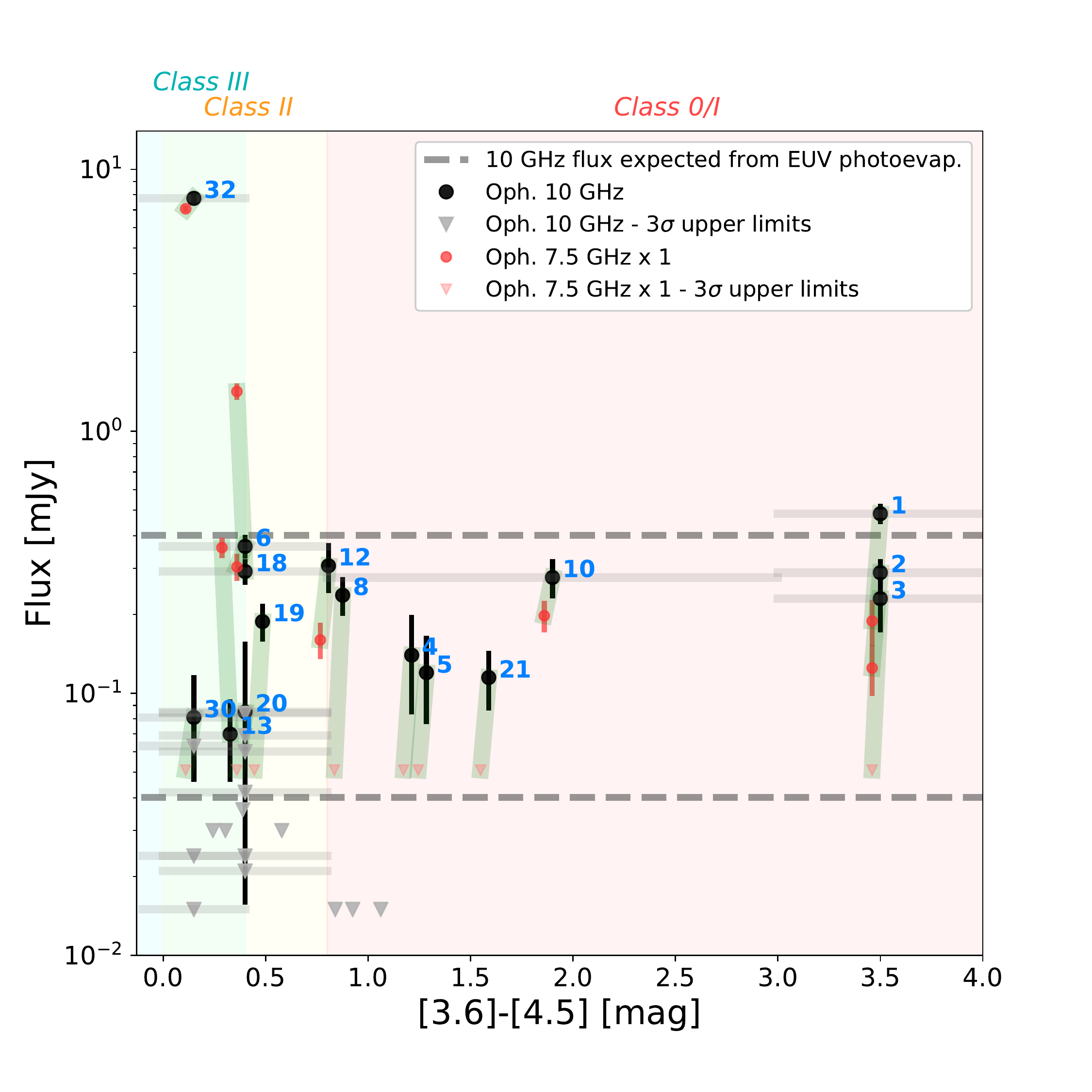} \\
   \end{tabular}
   \caption{Summary of the observed 10 GHz radio fluxes. We show the characteristic [3.6]-[4.5] color ranges of the Class III, II, and 0/I YSOs as blue, yellow, and red filled regions which are bounded in the horizontal axis by [-0.4, 0.4], [0.0, 0.8], and [0.8, 4.0], respectively (the overlapped area for Class III and II objects appears in green; see \citealt{Allen2004}). Our 10 GHz detections are presented as black circles. For sources that we detected at 10 GHz, we also present the fluxes measured at 7.5 GHz (red symbols) by \citet{Dzib2013}. For presentation purposes, we offset the [3.6]-[4.5] values of the red symbols by -0.04. The observations towards the same target sources are linked by green lines. Gray and red downward triangles are the 3$\sigma$ upper limits from these observations. Dashed lines show the expected radio fluxes from EUV photoevaporation winds from protoplanetary disks, assuming the EUV flux $\Phi_{\rm EUV}=$10$^{41}$ (bottom) and 10$^{42}$ photons s$^{-1}$ (top). 
   }
   \label{fig_radioflux}
\end{figure}

Figure \ref{fig_radioflux} shows the radio emission properties of the YSOs versus their Spitzer [3.6]-[4.5] colors \citep{Evans2009}.
We see that the measured fluxes at 10 GHz of some sources are significantly brighter than the fluxes measured at 7.5 GHz by \citet{Dzib2013}, while for other sources it is the opposite. The absence of systematic trend indicates that our data are probably not affected by flux calibration issues. 
We note that the classification of the continuous evolution of YSOs into Class 0/I, II, and III stages, taken from the literature, is to some extent artificial, and can be uncertain for YSOs that are transitioning from one stage to another. In addition, different catalogs or databases may report slightly different classifications, which are noted in Table \ref{table_catalog}.

\subsection{Nature of the emission at 10 GHz}
\label{sect_nature_continuum}

In this section, we evaluate how much of the flux measured towards the YSOs in our 10 GHz VLA observations is due to (i) thermal emission from dust, and (ii) other mechanisms such as free-free emission from ionized radio jets or photoevaporative winds, gyro-synchrotron emission from active magnetospheres, and synchrotron emission produced through the acceleration of cosmic-rays by jet or protostellar surface shocks (e.g., \citealt{Macias2016,Gibb1999,Forbrich2007,Padovani2016,Padovani2018}). 

\onecolumn
\begin{landscape}
\begin{longtable}{rlccccccc@{ }cl}
\caption{\label{table_catalog} Catalog of sources observed in the field of view of our observations grouped in categories.}  \\
\hline \hline
\# & Source$^{0}$ & Flux (mJy)$^{1}$ &  Flux (mJy)$^{1}$ & Flux (mJy)$^{1}$ & Flux (mJy)$^{1}$  & Source & Ref.$^3$  & Variable$^{4}$ &  $L_X$$^{5}$ & Other names \\ 
& (J2000 coordinates) & 10.0 GHz & 7.5 GHz & 4.5 GHz & 107 GHz & type$^{2}$ & & &  (10$^{29}$ erg s$^{-1}$) &  \\
 \hline
1 & J162626.31-242430.7 & 0.485 $\pm$ 0.033 & 0.189 $\pm$ 0.034 & 0.189 $\pm$ 0.034 & ~59.82 $\pm$ 0.47$^{1}$ & YSO 0?$^8$ & a & Y &  & VLA1623 B  \\ 
2 & J162626.39-242430.8 & 0.289 $\pm$ 0.030 & 0.125 $\pm$ 0.025 & 0.087 $\pm$ 0.030 & ~59.82 $\pm$ 0.47$^{1}$ &  YSO 0 & a & U &  & VLA1623 A \\ 
3 &J162627.83-242359.4 & 0.230$^7$ & $\lesssim$0.051 & $\lesssim$0.051 & 23.13 $\pm$ 0.46 & YSO/PC$^9$ & b & & & SM1 \\   
\hline
4 & J162617.23-242345.7 & 0.140$^7$ & $\lesssim$0.051 & $\lesssim$0.051 & 14.46 $\pm$ 0.29 & YSO I & c,d & & & A-MM33, CRBR12, \\
&&&&&&&&&&ISO-Oph 21\\
5 & J162621.36-242304.7 & 0.120$^7$ & $\lesssim$0.051 & $\lesssim$0.051 & ~~3.48 $\pm$ 0.67 & YSO I/0-I & c/d & & & GSS 30-IRS1, Elias 21 \\  
6 & J162621.72-242250.9 & 0.364 $\pm$ 0.030 & 0.304 $\pm$ 0.029 & 0.238 $\pm$ 0.017 & 30.71 $\pm$ 0.63 & YSO I/FS & a/c & Y$_{\rm 4.5 GHz}$$^{10}$ & & GSS 30-IRS3, LFAM1 \\ 
7 & ~\,162622.27-242407.1 & $\lesssim$ 0.015 & $\lesssim$0.051 & $\lesssim$0.051 & $\lesssim$0.16 & YSO I-FS  & c & & & CRBR25 \\ 
8 & J162623.58-242439.9 & 0.237 $\pm$ 0.035 & $<$ 0.06 & 0.125 $\pm$ 0.015 & $\lesssim$0.16 & YSO 0-I/FS & d/a,c & Y & 10.8 &  LFAM 3\\
9 & ~\,162625.49-242301.6  & $\lesssim$ 0.015 & $\lesssim$0.051 & $\lesssim$0.051 & ~~9.17 $\pm$ 0.44 & YSO I  & c & & & CRBR36 \\ 
10 & J162625.63-242429.4 & 0.277 $\pm$ 0.041 & 0.198 $\pm$ 0.023 & 0.218 $\pm$ 0.014 & 11.24 $\pm$ 0.44 & YSO I & a,d & Y$_{\rm 7.5 GHz}$$^{10}$ & & VLA1623 W, LFAM 4\\  
11 & ~\,162630.47-242257.1  & $\lesssim$ 0.021 & $\lesssim$0.051 & $\lesssim$0.051 & 11.22 $\pm$ 0.44 & YSO FS/0-I & c/d & & 4.8 & $\ldots$ \\  
\hline
12 & J162610.32-242054.9 & 0.307$^7$ & 0.160 $\pm$ 0.022 & 0.100 $\pm$ 0.012 & 27.25 $\pm$ 0.35 & YSO II & a,c,d & Y &  9.5 & GSS 26 \\ 
13 & J162616.85-242223.5 & 0.070$^7$ & 0.360 $\pm$ 0.024 & 0.337 $\pm$ 0.017 & $\lesssim$0.16 & YSO II & a,c,d & Y & 16.3 & GSS 29, LFAM p1 \\
14 & ~\,162617.06-242021.6 & $\lesssim$ 0.030 & $\lesssim$0.051 & $\lesssim$0.051 & $\lesssim$0.16 & YSO II & c & &  10.7 & DoAr 24  \\
15 & ~\,162618.82-242610.5 & $\lesssim$ 0.030 & $\lesssim$0.051 & $\lesssim$0.051 & $\lesssim$0.16 & YSO II & c & & & $\ldots$  \\
16 & ~\,162618.98-242414.3 & $\lesssim$ 0.015 & $\lesssim$0.051 & $\lesssim$0.051 & 5.96 $\pm$ 0.98 & YSO II-FS & c & & &  CRBR15 \\ 
17 & ~\,162621.53-242601.0 & $\lesssim$ 0.030 & $\lesssim$0.051 & $\lesssim$0.051 & $\ldots$ & YSO II & c & & 0.1--0.6 & $\ldots$  \\
18 & J162622.39-242253.4 & 0.292 $\pm$ 0.027 & 1.42 $\pm$ 0.07  & 2.02 $\pm$ 0.10 & $\lesssim$0.16& YSO II & a,c & Y &  51.5 & GSS 30-IRS2, VSSG12,  \\
&&&&&&&&&& ISO-Oph~34, LFAM 2 \\
19 & J162623.36-242059.9$^{6}$ & 0.188$^7$ & $\lesssim$0.051 & $\lesssim$0.051 & $\ldots$ & YSO II & c,d & &  4.7 & GSS 31a, DoAr 24Ea \\ 
20 & J162623.42-242102.0 & 0.085$^7$ & $\lesssim$0.051 & $\lesssim$0.051 & $\ldots$ & YSO II? & e & & & GSS 31b, DoAr 24Eb \\
21 & J162624.04-242448.5 & 0.115 $\pm$ 0.027 & $\lesssim$0.051 & $\lesssim$0.051 & $\lesssim$0.16 & YSO II/FS & c/d &  & 29.2 & S2 \\  
22 & J162625.28-242445.4$^{11}$ & $\lesssim$ 0.024 & $\lesssim$0.051 & $\lesssim$0.051 & $\lesssim$0.16 & YSO II & c &  & 0.5  & $\ldots$ \\
23 & ~\,162627.81-242641.8 & $\lesssim$ 0.036 & $\lesssim$0.051 & $\lesssim$0.051 & $\ldots$ & YSO II  & c & & & $\ldots$ \\
24 & ~\,162637.79-242300.7 & $\lesssim$ 0.042 & $\lesssim$0.051 & $\lesssim$0.051 & $\lesssim$0.16 & YSO II & c & & & LFAM p2 \\
25 & ~\,162642.74-242427.7 & $\lesssim$ 0.060 & $\lesssim$0.051 & $\lesssim$0.051 & $\lesssim$0.16 & YSO II  & c & & & $\ldots$  \\
26 & ~\,162642.89-242259.1  & $\lesssim$ 0.069 & $\lesssim$0.051 & $\lesssim$0.051 & $\lesssim$0.16 & YSO II  & c & & &$\ldots$  \\
27 & ~\,162643.86-242450.7 & $\lesssim$ 0.084 & $\lesssim$0.051 & $\lesssim$0.051 & $\lesssim$0.16 & YSO II  & c & & &$\ldots$  \\
\hline
28 & ~\,162615.81-241922.1 & $\lesssim$ 0.063 & $\lesssim$0.051 & $\lesssim$0.051 & $\lesssim$0.16 & YSO III & c & & 3.1 & $\ldots$  \\
29 & ~\,162622.19-242352.4 & $\lesssim$ 0.015 & $\lesssim$0.051 & $\lesssim$0.051 & $\lesssim$0.16 & YSO III  & c & & & $\ldots$  \\
30 & J162625.23-242324.3 & 0.081$^7$ &  $\lesssim$0.051 & $\lesssim$0.051 & $\lesssim$0.16 & YSO III/FS & c/d & & 6.0 &$\ldots$  \\
31 & ~\,162631.36-242530.2 & $\lesssim$ 0.024 & $\lesssim$0.051 & $\lesssim$0.051 & $\lesssim$0.16 & YSO III  & c & & 0.2 &$\ldots$  \\
32 & J162634.17-242328.7 &   7.75 $\pm$ 0.11 & 7.07 $\pm$ 0.35 & 7.98 $\pm$ 0.40 & $\lesssim$0.16 & YSO III & a,c & N & 22.6 & S1  \\
\hline 
33 & ~\,162614.63-242507.5 & $\lesssim$ 0.024 & $\lesssim$0.051 & $\lesssim$0.051 & $\lesssim$0.16  & YSO U &  f &  & & $\ldots$  \\ 
34 & ~\,162625.99-242340.5 & $\lesssim$ 0.015 & $\lesssim$0.051 & $\lesssim$0.051 & $\lesssim$0.16 &YSO U &  g & & & $\ldots$ \\
35 & ~\,162632.53-242635.4 & $\lesssim$ 0.045 & $\lesssim$0.051 & $\lesssim$0.051 & $\lesssim$0.16  & YSO U &  c & & & CRBR40 \\
36 & ~\,162638.80-242322.7 & $\lesssim$ 0.045 & $\lesssim$0.051 & $\lesssim$0.051 & $\ldots$ & YSO U & c & & & $\ldots$ \\
37 & ~\,162639.92-242233.4 & $\lesssim$ 0.078 & $\lesssim$0.051 & $\lesssim$0.051 & $\ldots$ & YSO U & c & & & $\ldots$ \\ 
\hline
38 & ~\,162608.04-242523.1 & $\lesssim$ 0.051 & $\leq$0.05 & 0.103 $\pm$ 0.014 &  & EG? & a & Y &  \\
39 & J162629.62-242317.3 &  0.091$^4$ & 0.124 $\pm$ 0.018 & 0.228 $\pm$ 0.014 & & EG? & a & Y &  \\
40 & J162630.59-242023.0 & $\lesssim$ 0.045 & 0.064 $\pm$ 0.017 & 0.098 $\pm$ 0.013 &  & EG? & a & Y & \\

\# & Source$^{0}$ & Flux (mJy)$^{1}$ &  Flux (mJy)$^{1}$ & Flux (mJy)$^{1}$ & Flux (mJy)$^{1}$  & Source & Ref.$^3$  & Variable$^{4}$ &  $L_X$$^{5}$ & Other names \\ 
& (J2000 coordinates) & 10.0 GHz & 7.5 GHz & 4.5 GHz & 107 GHz & type$^{2}$ & & &  (10$^{29}$ erg s$^{-1}$) &  \\
 \hline
41 & J162634.95-242655.3 & $\lesssim$ 0.069 & 0.100 $\pm$ 0.013 & 0.197 $\pm$ 0.019 &  & EG? & a & Y&     \\
42 & J162635.33-242405.3  & 0.377 $\pm$ 0.039 & 0.329 $\pm$ 0.033 & 0.650 $\pm$ 0.038 &  & EG? & a & Y &     \\ 
\hline
\end{longtable}
\tablefoot{ 
$^0$ The source name starts with J when detected with the VLA, either in this study or in \citet{Dzib2013}. The rest of the name correspond to the J2000 RA-Dec coordinates {\it hhmmss.ss-ddmmss.s}.  When undetected with VLA, we used the coordinates given in the references listed in column 8. \\
$^1$ The fluxes measured at 10.0 GHz were derived with Gaussian fit. The uncertainties correspond to the fit uncertainties only. The fluxes measured at 4.5 and 7.5 GHz come from \citet{Dzib2013}, while the ones at 107 GHz come from \citet{Kirk2017}. It should be noted that the flux measured at 107 GHz for J162626.31-242430.7 and J162626.39-242430.8 includes the two sources. The upper limits correspond to the 3$\sigma$ levels (1$\sigma$=17 $\mu$Jy beam$^{-1}$ for both frequencies) measured in the \citet{Dzib2013}'s 3-epoch combined images.\\
$^2$ The source is either YSO (Young Stellar Object) or EG (Extragalactic candidate). The YSOs are classified into: PC (Prestellar Core), 0 (Class 0 protostar), I (Class I protostar), FS (Flat Spectrum), II (Class II protostar), III (Class III protostar), U (Unknown classification for the YSO candidates). \\
$^3$ References for the YSO classification : a) \citet{Dzib2013}, b) \citet{Friesen2014}, c) \citet{Wilking2008}, d) \citet{Hsieh2013}, e) \citet{Kruger2012}, f) \citet{Jorgensen2008}, g) \citet{Evans2009}  \\
$^4$ Variability taken from Dzib et al. (2013). A source is considered variable when its variability fraction is $\geq$25\%. Legend: Y - variable; N - not variable; U - unknown. \\
$^{5}$ X-ray luminosity in the 0.5-9.0 keV \citep{Imanishi2003}. \\
$^6$ This source could be a binary. The flux given here corresponds to the total flux. \\
$^7$ Contrary to the other sources, the fluxes of these objects were integrated over a circular area selected with CASA due to their particular structure, which may be caused by residual phase errors, or due to a very uncertain Gaussian fit. \\
$^8$ This source may be a young star or an outflow knot feature. \\
$^9$ This source was proposed to contain an extremely young, deeply embedded protostellar object \citep{Friesen2014}.\\
$^{10}$ Source variable only at the indicated frequency. \\
$^{11}$ Source only detected in epoch 3 with a flux of 0.4 Jy. \\}
\end{landscape}
\twocolumn

\begin{figure*}
   \hspace{-0.8cm}   
   \begin{tabular}{ p{9cm} p{9cm} }
     \vspace{-0.4cm} \includegraphics[width=9.5cm]{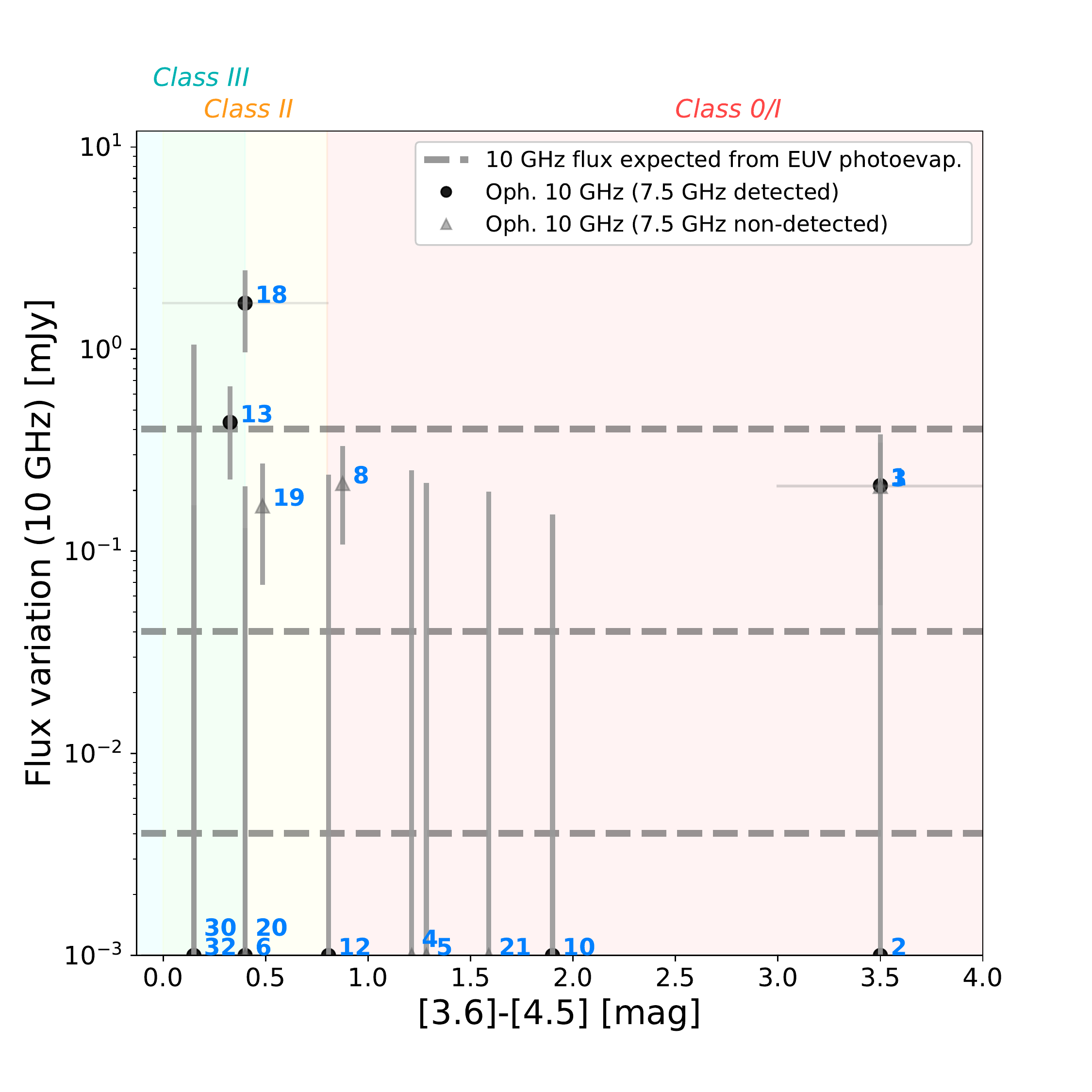} &
     \vspace{-0.4cm} \includegraphics[width=9.5cm]{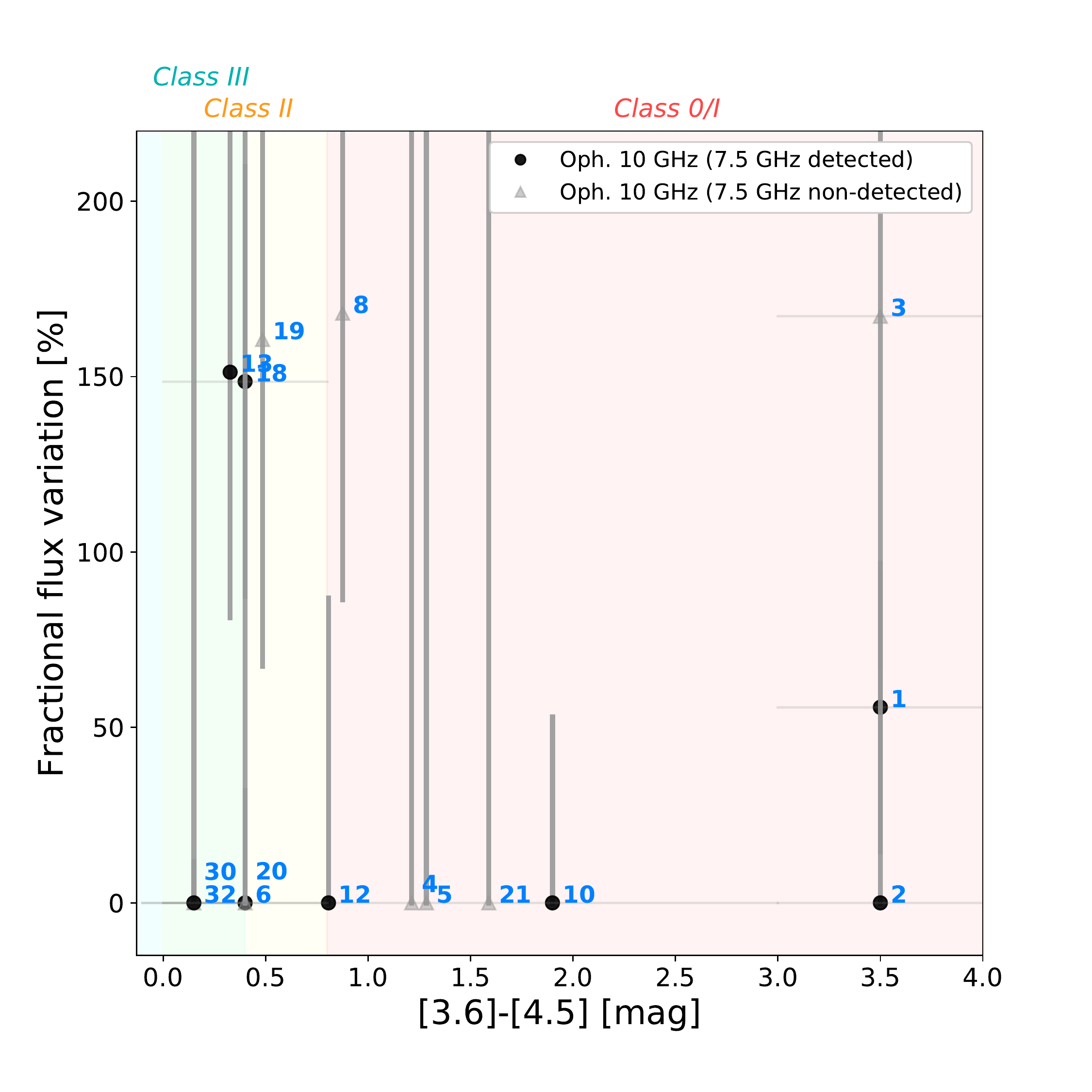} \\
   \end{tabular}
   \vspace{-0.65cm}
   \caption{
           Summary of the observed radio flux variability ({\it left panel}) and fractional radio flux variability ({\it right panel}). We omit sources which were not detected in both our 10 GHz observations and the previous 7.5 GHz observations of Dzib et al. (2013), since there is essentially no constraint on their time variability. For sources which were detected from at least one of those observations, we present the flux variation by calculating the average of the differences between the measured 10 GHz flux in our VLA observations and the expected 10 GHz flux derived by re-scaling the 7.5 GHz flux from \citet{Dzib2013} to 10 GHz assuming $\alpha=$-0.1 and 2.0. Vertical error bars take the measurement errors and the spectral index range [-0.1, 2.0] into consideration. 
           Dashed lines in the left panel show the expected radio fluxes from EUV photoevaporation winds from protoplanetary disks, assuming the EUV flux $\Phi_{\rm EUV}=$10$^{40}$ (bottom), 10$^{41}$ (middle), and 10$^{42}$ photons s$^{-1}$ (top). 
        }
   \label{fig_radioVar}
\end{figure*}

The brightest source in our sample, J162634.17-242328.7 (S1, \#32), has been already investigated in several studies and is known to be a completely non-thermal source. There is no evidence of a free-free component \citep[e.g., ][]{Andre1988,Loinard2008,Ortiz2017}. Indeed, the flux measured with the Very Long Baseline Array (VLBA) is systematically found to be equal to the VLA flux \citep{Loinard2008,Ortiz2017}. Since the VLBA is only sensitive to non-thermal emission whereas the VLA is in principle sensitive to both thermal and non-thermal emission \citep[e.g.,][]{Ortiz2017}, the emission of this source is confirmed here to be fully non-thermal. This source is, however, quite peculiar, since the non-thermal emission is not strongly variable, as confirmed with our observations (see Table \ref{table_catalog}). This result is somewhat of a mystery, and may be due to a magnetic field that, in this specific case, is fossil-based rather than dynamo driven \citep{Andre1988}. The former would explain S1's lack of flaring activity that is otherwise typically seen in non-thermal sources. Given these extended studies focused on the S1 source, we do not discuss it further.

\subsubsection{Contribution of the thermal emission from dust}

To determine the thermal contribution from dust, we assume that the $\sim$107 GHz continuum fluxes reported by \citet{Kirk2017} are entirely due to dust thermal emission, and then extrapolate the contribution of dust emission at our observing frequency of 10 GHz by assuming a power-law with a spectral index $\alpha$ (see Table \ref{table_catalog}). We note that the angular resolution of the observations reported by \citet{Kirk2017} is approximately 10 times coarser than that of our VLA observations.
Therefore, our estimates of 10 GHz dust emission should be regarded as upper limits.

In the millimeter bands (e.g., $\sim$90--350 GHz), the spectral indices of Class 0/I objects may be $\alpha$=2.5--3 (see \citealt{Chiang2012,Tobin2013,Tobin2015,Miotello2014}), while those of Class II/III objects may become lower ($\alpha$=2--2.5; \citealt{Ricci2010,Perez2012,Tazzari2016}) due to dust grain growth or high optical depths (see \citealt{Li2017,Galvan2018}).
Taking this difference into account, we find that dust thermal emission could account for up to $\sim$30\% of the observed 10 GHz flux toward the Class 0 YSOs and is almost negligible in the Class III objects of our sample.
For the Class I/II YSOs, the situation is more complex. In general, the contribution of the dust emission is $\leq$30\% and in some cases negligible.
Exceptions, however, include the Class II sources J162610.32-242054.9 (also known as GSS26, \#12), for which dust emission could account for $\sim$80\% of the continuum flux at 10 GHz, and  
162618.98-242414.3 (also called CRBR15, \#16), for which the predicted dust emission is higher than the upper limit of 15 $\mu$Jy beam$^{-1}$, as well as two Class I sources (162625.49-242301.6, \#9 and 162630.47-242257.1, \#11), for which the predicted dust emission fluxes are comparable to the measured upper limits of 15 $\mu$Jy beam$^{-1}$ at 10 GHz.
Therefore, except for a few Class I/II sources, the contribution from dust is in general $\leq$30\% of the total emission. This behavior is consistent with even higher-angular resolution 870 $\mu$m ALMA observations toward the Class II sources J162623.36-242059.9 (\#19) and J162623.42-242101.9 \citep[\#20,][]{Cox2017}, for which the dust contribution at 10 GHz is also estimated to be $\leq$30\% assuming dust spectral indices $\alpha$=2--2.5.  

\subsubsection{Nature of the remaining radio emission}
\label{sect_nature_remain_emission}

The remaining radio fluxes likely have contributions from (thermal) free-free emission from ionized radio jets, (thermal) free-free emission due to photoevaporative winds \citep[e.g.,][]{Macias2016} or (non-thermal) gyro-synchrotron emission from stellar magnetospheres \citep[e.g.,][]{Gibb1999,Forbrich2007}. Jet or protostellar surface shocks can also produce (non-thermal) synchrotron emission at our observing frequency, for example through the acceleration of cosmic-rays \citep{Carrasco2010,Padovani2016,Anglada2018}. These radio emission mechanisms present specific characteristics, which we describe below.

Free-free emission from thermal jets and (gyro-)synchrotron emission are known to be time variable \citep{Forbrich2007,Dzib2013}, but they may have very different characteristic timescales \citep{Liu2014}.
Free-free emission may vary on time-scales from a few weeks to a few months considering the ionized gas recombination timescales as well as the dynamical timescales of the inner $\sim$1 au disk. Gyro-synchrotron emission, however, is expected to vary on shorter timescales (minutes) due to flares on a stellar surface, and can vary up to the rotational periods of protostars. These periods can be as long as $\sim$10 days, due to large magnetic loops coupling protostars and their inner disks \citep{Forbrich2007,Liu2014}. 
Synchrotron emission is also expected to be variable, although the timescale is unclear \citep{Padovani2016}. 

Observations also indicate that the fluxes of thermal (free-free) sources vary rarely more than 20--30\%, while, in general, non-thermal sources show larger variability \citep{Ortiz2017,Tychoniec2018}.
The spectral indices of each type of emission can also differ. Free-free emission is characterized by spectral indices in the range [-0.1, 2.0], while gyro-synchrotron emission can span a significantly larger range of -5 to +2.5. Spectral indices $<$ -0.4 have been observed in YSO jets and attributed to synchrotron emission \citep{Anglada2018}.

To probe the origins of the detected emission, we first checked if any of our sources were also detected with the VLBA. As explained at the beginning of Section \ref{sect_nature_continuum}, any detection with the VLBA is necessarily non-thermal. In addition to S1, three other sources present in our observations:  J162616.85-242223.5 (GSS 29, \#13), J162622.39-242253.4 (GSS 30-IRS2, \#18), and J162625.63-242429.4 (VLA1623 W, \#10) are detected at 5 GHz with the VLBA \citep{Ortiz2017} but they are undetected at 8 GHz (see Table \ref{tab_fluxes_vlba}).
By comparing the VLBA fluxes to the VLA fluxes measured by \citet{Dzib2013}, we find that the emission of J162616.85-242223.5 (\#13) could be fully non-thermal at 5 GHz. Unfortunately, no flux is available at 8 GHz for this source and we cannot rule out a fully non-thermal emission at 10 GHz. The emission of J162622.39-242253.4 (\#18) could only be partially non-thermal, as the VLBA flux is lower than the VLA flux (19\% at 5 GHz and $<$6\% at 8 GHz). Nevertheless it has to be considered cautiously, since this source is possibly highly variable \citep{Dzib2013} and the observations were not carried out at similar dates.
The emission of J162625.63-242429.4 (\#10) could be fully non-thermal, since at 5 GHz the VLBA emission is higher than the VLA flux and the VLBA upper limit at 8 GHz is not even a factor of 2 lower than the VLA measurement.
  
\begin{table}[h!]
\begin{center}
\caption{Comparison of fluxes (mJy) measured towards 3 YSOs with the VLBA and the VLA. }
\begin{tabular}{@{ }r@{~~}c@{~~}c@{~~}c@{~~}c@{~~}c@{ }}
\hline \hline
\# & Source & VLBA$^b$   & VLA$^c$   & VLBA$^b$   & VLA$^c$   \\
& & 5 GHz & 5 GHz & 8 GHz & 8 GHz \\
\hline
10 & J162625.63-242429.4 & 0.66 & 0.22 & $<$0.12 & 0.20\\
13 & J162616.85-242223.5 & 0.15--0.47$^a$ & 0.34 & $\ldots$ & 0.36 \\
18 & J162622.39-242253.4 & 0.30--0.38$^a$ & 2.02 & $<$0.09 & 1.42  \\ 
\hline
\end{tabular}
\label{tab_fluxes_vlba}
\tablefoot{$^a$ The flux of these sources is known to vary. 
$^b$ From \citet{Ortiz2017}. $^c$ From \citet{Dzib2013}.}
\end{center}
\end{table}%

Next, we determined the spectral indices of all sources between 10 GHz and 7.5 GHz and between 10 GHz and 4.5 GHz, taking into account both the fit uncertainty and the calibration uncertainty (see column 3 in Table \ref{table_catalog_emission}).
The only two sources with negative spectral indices (J162616.85-242223.5, \#13 and J162622.39-242253.4, \#18) are the ones detected with the VLBA, which confirms the non-thermal origin of these sources' emission.
Four sources, J162626.31-242430.7 (\#1), J162627.83-242359.4 (\#3), J162623.58-242439.9 (\#8), and J162623.36-242059.9 (\#19), show spectral indices higher than 2.5, which may indicate variability (see below).

Finally, we explored the long-term variability of the YSOs. Our observations were averaged over a couple of months and compared with those of \citet{Dzib2013} obtained in 2011. \citet{Dzib2013} reported that seven out of the 16 YSOs we detected are variable (see Table \ref{table_catalog}, column 9). We note that among these variable sources, two are those with spectral indices between 7.5 GHz and 10 GHz higher than 2.5. They also include the sources of non-thermal emission detected with the VLBA. 

Any short-term variability will be explored in another paper by analyzing separately the 5 epochs as well as more recent observations at lower spatial resolution. Nevertheless, to ensure that our conclusions are not affected by significant variability, we separately checked the maps of the different epochs. As expected, the faintest sources are barely detected or not detected depending on the noise level of each epoch. Among the brightest sources, even if some variations are observed for some of them, the fluxes vary around the values measured in the map with the combined epochs. We did not see cases where the flux is significantly higher at one epoch. The only exception is found for the source J162625.8-242445.0 (\#22) that is not detected in the map with the combined epochs, but clearly detected in epoch 3 with a flux of 0.4 mJy (see Figure \ref{fig_indiv_maps}), which is probably due to a non-thermal flare. 

To explore possible radio flux variations since the observations of \citet{Dzib2013}, we extrapolated their fluxes at 7.5 GHz to those at 10 GHz assuming that $\alpha$ is in the range of [-0.1, 2.0] (i.e., free-free emission from optically thin to optically thick limits) and compared the resulting values to our measured fluxes. For sources which were not detected at 7.5 GHz by \citet{Dzib2013}, we evaluated the corresponding 3\,$\sigma$ limits at 10 GHz assuming $\alpha$=2.0, and compared the resulting values with our measurements (see Figure \ref{fig_radioVar}). We found that there are three sources (J162626.31-242430.7/\#1, J162616.85-242223.5/\#13, and J162622.39-242253.4/\#18) detected in both our 10 GHz observations and the previous 7.5 GHz observations, for which the flux differences are too large to be explained by constant free-free emission. The emission of J162616.85-242223.5 (\#13), and J162622.39-242253.4 (\#18) is certainly non-thermal, as explained before. The emission of J162626.31-242430.7 (\#1) may be explained either by non-thermal radio emission or by thermal radio flux variability of more than a few tens of percent (see Figure \ref{fig_radioVar}). 
In addition, after considering the spectral index range [-0.1, 2.0], it appears that three of our new radio detections (J162627.83-242359.4/\#3, J162623.58-242439.9/\#8, and J162623.36-242059.9/\#19) cannot be attributed to our improved  sensitivity. The measured 10 GHz fluxes in the new VLA observations are significantly larger than 10 GHz fluxes scaled from the 7.5 GHz upper limit fluxes of \citet[][see Figure \ref{fig_radioVar}]{Dzib2013}.
Therefore, these detections are either due to variability or  non-thermal, gyro-synchrotron spectral indices. 
The fractional radio flux variability of the sources can be seen in Figure \ref{fig_radioVar}. We find that six out of our detected sources in the [3.6]-[4.5] color range of [0, 2] (i.e., late Class 0/I to early Class III stages) show over 50\% fractional radio flux variability. The absolute values of their flux variations appear comparable to the observed flux variations from five epochs of observations towards CrA on the same date \citep{Liu2014}.
The radio emission of some of these six sources (including J162625.63-242429.4/\#10, 162616.85-242223.5/\#13, and J162622.39-242253.4/\#18) may be largely contributed by gyro-synchrotron emission which can vary on short timescales. 

Table \ref{table_catalog_emission} summarizes our conclusions regarding the radio emission of the Oph A YSOs.

\begin{table*}[!h]
\begin{center}
\caption{Summary of the emission of the YSOs.} 
\label{table_catalog_emission} 
\begin{tabular}{@{}r c c c c c c@{ }c@{ }c@{}}
\hline \hline
\# & Source & Spectral & Dust & \multicolumn{3}{c}{Ionized emission} & X-ray &  $\phi$$_{EUV}$$^{6}$\\
\cline{5-7}
 & &  index & & VLBA & Variability$^3$ &  Fully or    &  detection$^5$ & (10$^{40}$ \\ 
& &    $\alpha$$^1$   & & detection  &  & partially  &  & erg s$^{-1}$)\\
& &    &    & at 5 GHz$^2$ &         &  non-thermal  \\
& &    &    &  & & emission$^4$ & \\
 \hline
 \multicolumn{9}{c}{Class 0} \\
  \hline
1 & J162626.31-242430.7 & 3.3$\pm$0.7 / 1.2$\pm$0.3 & $\leq$ 21\% & N & Y  & Y  & \\
2 & J162626.39-242430.8 & 2.9$\pm$0.8  / 1.5$\pm$0.5 & $\leq$ 21\% & N &  \\ 
3 & J162627.83-242359.3 & $\geq$4.8  /  $\geq$1.7 & $\leq$ 27\% & & Y & Y \\ 
\hline
 \multicolumn{9}{c}{Class I} \\
 \hline
4 & J162617.24-242346.0 & $\geq$2.9 /  $\geq$1.7 & $\leq$ 28\% & &  \\
5 & J162621.36-242304.7 & $\geq$2.3  /  $\geq$1.0 & $\leq$ 8\% & &  \\ 
6 & J162621.72-242250.9 & 0.6$\pm$0.5  /  0.5$\pm$0.2 & $\leq$ 23\% & N & Y$_{\rm 4.5GHz}$$^\dagger$ & Y \\ 
7 & ~162622.27-242407.1 & $\ldots$  & $\ldots$  \\ 
8 & J162623.58-242439.9 &  $>$4.2  /  0.8$\pm$0.3  & $\leq$ 0.2\% & N & Y  & Y & Y  \\
9 & ~162625.49-242301.6 & $\ldots$  & $\leq$ 100\% \\ 
10 & J162625.63-242429.4 & 1.2$\pm$0.7  /  0.3$\pm$0.2 & $\leq$ 10\% & Y & Y$_{\rm 7.5GHz}$$^\dagger$ & Y \\ 
11 & ~162630.47-242257.1  & $\ldots$  & $\leq$ 100\% & & & & Y\\ 
\hline
 \multicolumn{9}{c}{Class II} \\
  \hline
12 & J162610.32-242054.9 &  2.3$\pm$0.6  / 1.4$\pm$0.2 & $\leq$ 77\% & N & Y$^\dagger$ & Y &  Y &  $<$74 \\ 
13 & J162616.85-242223.5 & -5.7$\pm$1.0 / -1.9$\pm$0.4 & $\leq$ 2\% & Y & Y  & Y & Y & \\ %
14 & ~162617.06-242021.6 & $\ldots$  & $\ldots$ & & & & Y & $\lesssim$7\\
15 & ~162618.82-242610.5 & $\ldots$  & $\ldots$ & & & &   & $\lesssim$7 \\
16 & ~162618.98-242414.3 & $\ldots$  & $\leq$ 100\% & & & &   & $\lesssim$4 \\ 
17 & ~162621.53-242601.0 & $\ldots$  & $\ldots$ & & & & Y & $\lesssim$7\\
18 & J162622.39-242253.4 & -5.5$\pm$0.4   /  -2.4$\pm$0.2 & $\leq$ 0.5\% & Y & Y  & Y & Y & $<$71 \\ 
19 & J162623.36-242059.9 & $\geq$4.0  /  $\geq$1.4 & $\leq$ 18\% & & Y & Y  & Y & $<$45\\
20 & J162623.42-242101.9 & $\geq$0.8  /  $\geq$0.3 & $\leq$ 34\% & & & & & $\lesssim$21 \\
21 & J162624.04-242448.5 & $\geq$1.9 / $\geq$0.7 & $\leq$ 1\% & &  & & Y & $\lesssim$28 \\ 
22 & J162625.28-242445.4 & $\ldots$  & $\ldots$ & & & & Y & $\lesssim$6 \\
23 & ~162627.81-242641.8 & $\ldots$  & $\ldots$ & & & &   & $\lesssim$9 \\
24 & ~162637.79-242300.7 & $\ldots$  & $\ldots$ & & & &   & $\lesssim$10 \\
25 & ~162642.74-242427.7 & $\ldots$  & $\ldots$ & & & &   & $\lesssim$14 \\
26 & ~162642.89-242259.1 & $\ldots$  & $\ldots$ & & & &   & $\lesssim$17 \\
27 & ~162643.86-242450.7 & $\ldots$  &  $\ldots$ & & & &   & $\lesssim$21 \\ 
\hline
 \multicolumn{9}{c}{Class III} \\
  \hline
28 & ~162615.81-241922.1 & $\ldots$  & $\ldots$  & & & & Y & $\lesssim$15 \\
29 & ~162622.19-242352.4 & $\ldots$  &  $\ldots$  & & & &   & $\lesssim$4 \\
30 & J162625.23-242324.3 & $\geq$0.3 /  $\geq$0.1 & $\leq$ 2\%  & & & & Y & $\lesssim$20 \\
31 & ~162631.36-242530.2 & $\ldots$  &  $\ldots$ & & & & Y & $\lesssim$6 \\
32 & J162634.17-242328.7 & 0.3$\pm$0.3  /  0.0$\pm$0.1 & $\leq$ 0.02\%  & Y & N & Y & Y  &  \\ 
\hline
\end{tabular}
\tablefoot{
$^1$The first spectral index is calculated between 7.5 and 10 GHz, the second between 4.5 and 10 GHz. The uncertainties take into account both the fit uncertainty and the calibration uncertainty. The spectral indices can be significantly affected by variability, as the measurements were carried out at several epochs. 
$^2$\citet{Ortiz2017}. Y for detection, N for non detection.
$^3$\citet{Dzib2013} and this study. The sources with the symbol $^\dagger$ correspond to the sources that only present variability ($>$ 25\%) in \citet{Dzib2013}.
$^4$ Y for the sources expected to present non-thermal emission based on the criteria discussed in the text (VLBA detection, variability, and spectral indices). Some of these sources could also be sources with thermal emission but abnormally high variability.
$^5$\citet{Imanishi2003}. 
$^{6}$EUV luminosities reaching the disks calculated from the fluxes or 3$\sigma$ noise levels measured in our VLA images at 10 GHz (3$\,$cm) and using Eq. (2) of \citet{Pascucci2012} and a distance of 137 pc for the Oph A cluster.  \\ }
\end{center}
\end{table*}

\subsubsection{Association with X-ray emission}

We checked the sources associated with X-ray emission \citep[][see column 10 in Table \ref{table_catalog}]{Imanishi2003}. For Class III sources, X-ray emission mainly arises from magnetized stellar coronae, while in younger (Class I/II) sources, additional mechanisms can produce X-ray emission (e.g., shocks due to the material infalling from the disk to the stellar surface or due to the interaction of outflows with circumstellar material). All the Class II and III sources detected in our data are associated with X-ray emission, apart from J162623.42-242102.0 (DoAr 24Eb, \#20). The spatial resolution of Chandra telescope might not be sufficient to separate its emission from J162623.36-242059.9 (DoAr 24Ea). Among the younger sources we detected, only the Class I object J162623.58-242439.9 (\#8) is detected in X-ray. 

\section{Discussion: Revisiting photoevaporation in Class II/III proto-planetary disks}
\label{sect_discussion}

High energy stellar photons (UV or X-rays) may contribute to the dispersal of protoplanetary disks through photoevaporation \citep{Hollenbach1994, Alexander2014}. 
The exact contribution of this mechanism to disk dispersal and the way it impacts planet formation, however, need to be further investigated. 
Observations at radio wavelengths can probe the free-free emission from a disk surface that is partially or totally ionized by EUV photons or X-ray photons. Therefore, radio wavelength observations can be a powerful diagnostic of the contributions of these two types of photons in the protoplanetary disk evolution. 
For example, \citet{Pascucci2012} predict the level of radio emission expected from photoevaporation driven by EUV photons or X-ray photons. 
Based on an analysis of 14 circumstellar disks, \citet{Pascucci2014} then determined that the EUV photoevaporation mechanism may not play a significant role in disk mass dispersal, when EUV photon luminosities ($\Phi$ $_{\rm EUV}$) are lower than 10$^{42}$ photons s$^{-1}$.  
Similar conclusions were obtained by \citet{Galvan2014} for ten disks toward the Corona Australis (CrA) star-forming region, inferring $\Phi_{\rm EUV}$ < (1--4) $\times$ 10$^{41}$ photons s$^{-1}$, and by \citet{Macias2016} for the transitional disk of GM Aur ($\Phi_{\rm EUV}$ $\sim$ 6 $\times$ 10$^{40}$ photons s$^{-1}$).

\subsection{Constraints on EUV disk photoevaporation}

The high sensitivity of our observations (5 $\mu$Jy beam$^{-1}$ at the center of the field of view) and the proximity of this cloud (137 pc) allow us to derive stringent constraints on the contribution of EUV photons on disk photoevaporation in the Oph A star-forming region. As explained before, the radio emission of five of our detected Class II/III sources (J162610.32-242054.9/\#12, J162616.85-242223.5/\#13, J162622.39-242253.4/\#18, J162623.36-242059.9/\#19, and J162634.17-242328.7/\#32) is probably fully or partially non-thermal and we cannot exclude it for the three other detected sources. As such, the best constraints come from the Class II/III objects we did not detect.

Following the approach of \citet{Pascucci2014} and \citet{Galvan2014}, we estimate the expected radio continuum fluxes $F_{\mbox{\tiny 10 GHz}}$ for a particular EUV luminosity $\Phi_{\mbox{\tiny EUV}}$ based on the following formulation :
\begin{equation}
F_{\mbox{\tiny 10 GHz}} \mbox{ [$\mu$Jy]} \sim 4.0\times10^{-40}\left(\frac{137}{d \mbox{ [pc]}}\right)^{2} \left( \Phi_{\mbox{\tiny EUV}} \mbox{ [s$^{-1}$]} \right) \left(\frac{10.0}{8.5} \right)^{\alpha},
\end{equation}
where $d$ is the distance of the target source, and $\alpha$ is the spectral index of the free-free emission produced by the EUV photoevaporation.
As an approximation, we tentatively consider $\alpha$=0, and note that our estimate of $F_{\mbox{\tiny 10 GHz}}$ is not especially sensitive to the exact value of $\alpha$ as long as $\alpha$ is in the range of [-0.1, 2.0].
We provide the estimates of $F_{\mbox{\tiny 10 GHz}}$ at $\Phi_{\rm EUV}=$10$^{40}$, 10$^{41}$, and 10$^{42}$ photons s$^{-1}$ for Figures \ref{fig_radioflux} and \ref{fig_radioVar}. 
For Class II and III sources which were not detected in our  observations, the respective 3\,$\sigma$ upper limits of $F_{\mbox{\tiny 10 GHz}}$ constrain their $\Phi_{\rm EUV}$ to be $\lesssim$ 4--21 $\times$ 10$^{40}$ photons s$^{-1}$ (Figure \ref{fig_radioflux} and Table \ref{table_catalog}). 
These upper limits are lower than those derived from previous observations towards CrA \citep[<1--4 $\times$ 10$^{41}$ photons s$^{-1}$,][]{Galvan2014}. 
We note that typical EUV photoevaporation models require $\Phi_{\rm EUV}$ to be in the range of 10$^{41}$--10$^{42}$ s$^{-1}$ to disperse protoplanetary disks within a few Myrs \citep{Font2004,Alexander2006,Alexander2009}. EUV-driven photoevaporation is consequently very unlikely to play a major role in the dispersal of these disks.

For the Class II and III sources that are detected in our 10 GHz observations and do not necessarily exhibit non-thermal emission (J162623.42-242102.0/\#20, J162624.04-242448.5/\#21, and J162625.23-242324.3/\#30), if we assume that their 10 GHz fluxes are dominated by photoevaporation winds, the corresponding $\Phi_{\rm EUV}$ values are well in the range required by the aforementioned models (Figure \ref{fig_radioflux}). Hence, photoevaporation driven by EUV photons could be sufficiently efficient to disperse these disks.
Presently, however, we do not have good enough constraints about the spectral indices of these detected sources to tell what fractions of their radio fluxes come from constant EUV photoevaporation winds. 
Observationally, we also do not know yet whether the radio emission associated with EUV photo-evaporating disks evolves with time. 

\subsection{Constraints on X-ray disk photo-evaporation}

Photoevaporation by X-ray photons is another process that may lead to the dispersal of protoplanetary disks. We listed in Table \ref{table_catalog} the observed X-ray luminosities found in the literature for the YSOs of Oph A \citep{Imanishi2003}. They range over 0.01--3 $\times$ 10$^{30}$ erg s$^{-1}$.
\citet{Pascucci2012} determined the relation between the incident X-ray photon luminosity $L_{\mbox{\tiny X}}$ and the resulting free-free emission that a disk would emit:
\begin{equation}
F_{\mbox{\tiny 10 GHz}} \mbox{ [$\mu$Jy]} \sim 3.3\times10^{-30}\left(\frac{137}{d \mbox{ [pc]}}\right)^{2} \left(L_{\mbox{\tiny X}} \mbox{ [erg s$^{-1}$]} \right) \left(\frac{10.0}{8.5} \right)^{\alpha}.
\end{equation}
Based on this equation and the level of non-detections in our Class II objects, the upper limits derived for the incident X-ray photon luminosity are $\lesssim$ (7--25) $\times$ 10$^{30}$ erg s$^{-1}$, i.e. about 1--2 orders of magnitude higher than the observed values on average. 
Thus, we cannot exclude, with the present data, X-ray  photoevaporation as a major mechanism in the dispersal of the disks. More sensitive observations are needed to determine its efficiency.

\subsection{Studying the photoevaporation of protoplanetary disks with the Square Kilometre Array}

In the future, SKA will certainly revolutionize our understanding of the star and planet formation process through radio emission studies. We discuss here the potential of SKA to investigate the photoevaporation of disks.

The free-free emission produced by a disk (at the distance of Oph A) with an X ray luminosity of more than 10$^{29}$ erg s$^{-1}$ could be detected for example with an rms of 0.1 $\mu$Jy. Such a high sensitivity should be reached in the future with the SKA. In particular, \citet{Hoare2015} estimated that a 1000 hour deep field integration at the full resolution of SKA1-Mid ($\sim$40 mas, i.e. $\sim$5 AU for the disks of Oph A) over a 2 $\times$ 2.5 GHz bandwidth from 8.8 GHz to 13.8 GHz will yield a noise level of 0.07 $\mu$Jy beam$^{-1}$. 
Although the required amount of time appears significantly higher than the time dedicated to current radio projects, it should be noted that multiple projects will be carried out simultaneously with the SKA and that a large number of sources will be covered in the same field with a single pointing. For example, the investigation of the  photoevaporation in disk dispersal can be carried out simultaneously with the high priority studies of grain growth and the search for prebiotic molecules \citep{Hoare2015}. 

With a single pointing, the SKA will cover a field of view of about 6 arcminutes (comparable to our 4 pointing VLA mosaic). 
By targeting a rich region such as the Oph A cluster, a large number of disks (all the disks listed in this paper) can be observed simultaneously. 

For bright radio emission sources, SKA will further provide good constraints on the instantaneous spectral indices over a wide range of frequency, useful data for gauging the fractional contributions of thermal and non-thermal emission mechanisms. An expansion of SKA1-Mid to $\sim$25 GHz would provide even stronger constraints on the spectral indices resolved across the young stars, spatially separating the different components. Complementary observations will also be possible with the next generation VLA (ng-VLA, \citealt{Murphy2018,Selina2018}) above the highest SKA1-Mid band.

In addition, shallow (e.g., RMS $\sim$ few $\mu$Jy) but regularly scheduled SKA monitoring surveys will provide for the first time the statistics of how much time Class 0-III YSOs are in the radio active or inactive states, and what the dominant radio emission mechanisms and radio flux variability levels are during these states.

Finally, the SKA1-Mid resolution will be around 40 mas and hence it will be possible to separate spatially the different contributions from flares, jet, wind and disk to some degree.
Simultaneous observations of hydrogen radio recombination lines at the high-angular resolution of the SKA will also enable the separation of ionized gas emission from dust emission in disks, which will be key for these kinds of studies.

Getting photoevaporation rates should consequently be achievable with the power of SKA, however separating out the role of each type (EUV/X-ray) may be more complicated. According to \citet{Pascucci2012}, the EUV contribution should be a factor 10 higher than the X-ray contribution. Photoevaporation models predict different mass-loss profiles, but the subtraction of the EUV contribution to the free-free emission (necessary to investigate the X-ray driven photoevaporation of disks) could turn out to be highly uncertain, since the EUV luminosity is unknown.

\section{Conclusions}

We carried out very sensitive continuum observations of the Oph A star-forming region at 10 GHz with the VLA (1 $\sigma$ = 5 $\mu$Jy beam$^{-1}$ at the center of the field of view). We detected sixteen YSOs and two extragalactic candidate sources. Seven of the detected YSOs were not detected in a previous VLA survey of this region at 4.5 GHz and 7.5 GHz by \citet{Dzib2013}.

Using typical spectral indices for the possible components of radio emission, we constrained the origin of the emission detected at 10 GHz towards the YSOs. In general, dust emission contributes less than 30\% of the total emission. The 10 GHz emission appears to be mainly due to gyro-synchrotron emission from active magnetospheres, free-free emission from thermal jets or photoevaporative winds, or synchrotron emission due to accelerated cosmic-rays. Three of the YSOs show evidence of non-thermal emission. A comparison with the survey by \citet{Dzib2013} shows that six of the sources show over 50\% fractional radio flux variability, which is probably due to non-thermal emission. 

Constraints on the EUV and X-ray photoevaporation mechanisms were discussed. For the Class II/III disks for which we detect no emission, the corresponding EUV luminosities are not sufficient to explain disk dispersal within a few Myrs through theoretical photoevaporation models. For the sources detected at 10 GHz (with a possibly significant contribution of ionized thermal emission), the corresponding maximum $\Phi_{\rm EUV}$ values are within the range predicted by models. It is, however, currently unclear if EUV photoevaporating winds and their contributions to radio fluxes are constant in time.  
Even with the very high sensitivity of our observations, we are unable to provide strong constraints on the efficiency of X-ray for disk dispersal. Significantly more sensitive observations that also resolve the sources are necessary to locate the different emission origins and constrain the efficiency of the photoevaporation mechanisms.
With higher sensitivity and higher angular resolution, future facilities such as the Square Kilometre Array will make this possible.

\begin{acknowledgements}
This collaboration arose from discussions within the ``Cradle of Life'' Science Working Group of the SKA. The authors thank Hsieh Tien-Hao for providing the results of the classification method presented in \citet{Hsieh2013}. The National Radio Astronomy Observatory is a facility of the National Science Foundation operated under cooperative agreement by Associated Universities. A.C. postdoctoral grant is funded by the ERC Starting Grant 3DICE (grant agreement 336474). I.J.-S. acknowledges the financial support received from the STFC through an Ernest Rutherford Fellowship (proposal number ST/L004801). L.L. acknowledges the financial support of DGAPA, UNAM (project IN112417), and CONACyT, M{\'e}xico.
A.C.T. acknowledges the financial support of the European Research Council (ERC; project PALs 320620).  D.J. is supported by the National Research Council Canada and by an NSERC Discovery Grant. L.M.P. acknowledges support from CONICYT project Basal AFB-170002 and from FONDECYT Iniciaci\'on project \#11181068. A.P. acknowledges the support of the Russian Science Foundation project 18-12-00351. D.S. acknowledges support by the Deutsche Forschungsgemeinschaft through SPP 1833:
``Building a Habitable Earth'' (SE 1962/6-1). M.T. has been supported by the DISCSIM project, grant agreement 341137 funded by the European Research Council under ERC-2013-ADG. C.W. acknowledges 
support from the University of Leeds and the Science and Technology 
Facilities Council under grant number ST/R000549/1. This work was partly supported by the Italian Ministero dell'Istruzione, Universit\`a e Ricerca through the grant Progetti Premiali 2012 – iALMA (CUP C52I13000140001), by the Deutsche Forschungs-gemeinschaft (DFG, German Research Foundation) - Ref no. FOR 2634/1 TE 1024/1-1, and by the DFG cluster of excellence Origin and Structure of the Universe (\url{www.universe-cluster.de}). This project has received funding from the European Union’s Horizon 2020 research and innovation programme under the Marie Skłodowska-Curie grant agreement No 823823. This project has also been supported by the PRIN-INAF 2016 ``The Cradle of Life - GENESIS-SKA (General Conditions in Early Planetary Systems for the rise of life with SKA)''. 
\end{acknowledgements}

\bibliographystyle{aa} 
\bibliography{Biblio}

\begin{appendix}

\section{Image component sizes obtained with imfit.} 

\begin{table}[!h]
\caption{Image component sizes (deconvolved from beam) obtained with imfit.} 
\label{table_imfit} 
\begin{tabular}{@{}rccccc@{}}
\hline \hline
\# & Source & Major axis  & Minor axis  & Position  \\ 
& & FWHM ($\arcsec$)  & FWHM ($\arcsec$) & angle ($\degree$) \\
 \hline
1 & J162626.31-242430.7 & 0.78$\pm$0.09 & 0.52$\pm$0.09 & 76$\pm$15  \\
2 & J162626.39-242430.8 & 0.83$\pm$0.14 & 0.45$\pm$0.11 & 67$\pm$16 \\ 
\hline
6 & J162621.72-242250.9 & 0.97$\pm$0.11 & 0.57$\pm$0.05 & 37$\pm$8 \\ 
8 & J162623.58-242439.9 &  0.94$\pm$0.20  & 0.75$\pm$0.14 & 36$\pm$77 \\
10 & J162625.63-242429.4 & 1.06$\pm$0.22 & 0.69$\pm$0.23 & 85$\pm$28 \\ 
\hline
18 & J162622.39-242253.4 & 1.08$\pm$0.14 & 0.52$\pm$0.05 & 26$\pm$6 \\
21 & J162624.04-242448.5 & 1.37$\pm$0.40 & 0.47$\pm$0.15 & 52$\pm$12 \\ 
\hline
32 & J162634.17-242328.7 & 0.65$\pm$0.02 & 0.09$\pm$0.01 & 25.6$\pm$0.5 \\ 
\hline 
42 & J162635.33-242405.3  & 0.55$\pm$0.14 & 0.25$\pm$0.07 & 50 $\pm$16   \\ 
\hline
\end{tabular}
\tablefoot{The sources are listed in the same order as in Table \ref{table_catalog}.}
\end{table}

\end{appendix}

\end{document}